\documentclass[prb,aps,twocolumn,showpacs]{revtex4}
\usepackage{eurosym}
\usepackage{graphicx,color,psfrag,latexsym}
\usepackage{dcolumn}
\usepackage{amsmath,amssymb,epsf,bm}

\begin{document}

\title{AC Josephson Effect Induced by Spin Injection}
\author{A.~G. Mal'shukov$^{1}$ and Arne Brataas$^{2}$}
\affiliation{$^1$Institute of Spectroscopy, Russian Academy of Sciences,
142190, Troitsk, Moscow oblast, Russia \\
$^2$Department of Physics, Norwegian University of Science and Technology,
N-7491 Trondheim, Norway}

\begin{abstract}
Pure spin currents can be injected and detected in conductors via ferromagnetic contacts.
We consider the case when the conductors become superconducting. A DC pure spin current flowing in one superconducting wire towards
another superconductor via a ferromagnet contact induces AC voltage oscillations caused
by Josephson tunneling of condensate electrons. Quasiparticles
simultaneously counterflow resulting in zero total electric current through
the contact. The Josephson oscillations can be accompanied by
Carlson-Goldman collective modes leading to a resonance in the voltage
oscillation amplitude.
\end{abstract}

\maketitle

\section{Introduction}
\bigskip Electric and spin transport near ferromagnet-paramagnet interfaces
received a large attention boost with the discovery of the giant
magnetoresistance effect \cite{Fert:rmp2008} and the subsequent
developments in magnetoelectronics and spintronics. Aronov
\cite{Aronov}, and later Johnson and Silsbee \cite{Johnson}
theoretically predicted that an electric current through such an
interface leads to an accumulation of nonequilibrium spin
polarization with an accompanying spin current in the paramagnetic
metal. A reverse effect also takes place, a pure spin current from
the normal metal gives rise to an electric potential difference in
the ferromagnet. The physics of these phenomena is quite simple. A
sufficiently large difference of conductivities of spin-up and
spin-down electrons in ferromagnets induces spin polarization of the
electric current therein. Spin polarized currents passing through
ferromagnet-paramagnet boundaries result in an accumulation of
nonequilibrium magnetization near the interface. Both the spin
injection and detection of this spin polarization has been
experimentally demonstrated in Refs.
\onlinecite{Johnson2,Wees,Wees2} in systems containing two or more
junctions of thin normal metal wires with ferromagnets. One of them
acts as a spin injector, while the other is a detector, where the
voltage created by diffusing spins can be measured. Related
spin-polarized transport phenomena have been investigated in many
spintronic applications, such as giant magnetoresistance
\cite{Fert:rmp2008}, spin Hall effects \cite{Engel:2008}, current
induced magnetization dynamics \cite{Ralph:jmmm2008}, spin-pumping
\cite{Tserkovnyak:rmp2005}, and spin caloritronics
\cite{Bauer:ssc2010}.

In the case of superconducting systems, spin injection and detection
within a nonlocal setup similar to the one studied in Ref.
\onlinecite{Johnson2,Wees,Wees2} was investigated both theoretically
\cite{Zhao,Takahashi,Morten} and experimentally \cite{experim}.
These studies have been focused on DC transport. They revealed a
strong renormalization of spin-related transport parameters as
compared to normal systems. These changes were mostly caused by the
modified density of states in a superconductor. Beyond such
quasi-particle transport properties, the macroscopic coherent state
of the superconducting condensate can give rise to a quite different
transport phenomenon associated with the spin-polarized transport.
\begin{figure}[bp]
\includegraphics[width=8cm]{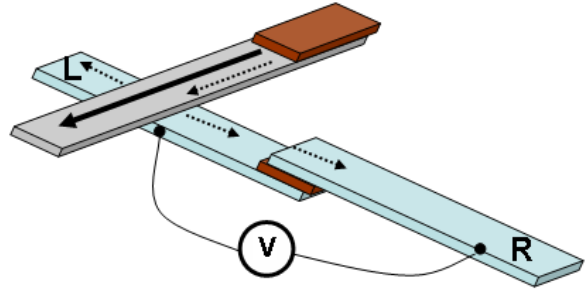}
\caption{A possible setup for observation of Josephson voltage
oscillations. A spin current (dashed arrows) is injected by passing
a DC current (solid arrow) from the ferromagnetic contact (brown).
The DC spin current through the ferromagnetic contact between the
right (R) and left (L) superconducting electrodes  induces periodic
oscillations of their electric potential difference $V$.}
\label{fig1}
\end{figure}

Below we will consider an AC effect produced by a DC spin current
towards a thin ferromagnetic contact. The DC potential induced by
this polarization flux gives rise to an AC electric current of
condensate electrons. Since in the considered experimental setup the
total current of the superconducting and normal components must be
zero, the AC condensate oscillations result in an AC potential
difference between the opposite sides of the contact. A schematic of
a possible experimental setup is shown in Fig. 1. A current is
passed from a ferromagnet to a normal metal generating an associated
spin accumulation and spin current therein. In the non-local
geometry, this spin accumulation also diffuses transversely in a
contacted normal metal towards another normal metal reservoir via a
ferromagnet contact. The non-local potential $V$ increases with the
injected DC current $I$ and the nonlocal resistance $R_{nl}=V/I$
describes the spin transport properties in the device. We will
demonstrate that when the normal metals become superconducting,
$R_{nl}$  acquires an AC component in addition to the DC component
in the normal state.

The outline of this paper is as follows. A model system used in our
calculation is described in Sec. III. Also in this section we
present a simple calculation of the AC voltage oscillations assuming
a local thermodynamic equilibrium between quasiparticles and the
condensate. A microscopic analysis based on coupled kinetic
equations for the superconducting order parameter and the
quasiparticle distribution function will be given in Sec. III. A
discussion of results will be presented in Sec. IV.

\section{AC voltage oscillations in a local thermodynamic equilibrium}

Our  model system consists of two superconducting wires in contact
via a spin-active barrier. We consider this contact to be weak in
the form of a thin ferromagnetic layer with, if necessary,
additional insulating layers. Such a
barrier can be characterized by two resistances $R_{\uparrow }$ and $%
R_{\downarrow }$ corresponding to two spin eigenstates. We assume
that a non-equilibrium spin polarization is created in the left wire
(see Fig.1), either by spin injection, as shown in Fig. 1, or by
other means. Moreover, we assume that the electron's energy
relaxation is faster than their spin relaxation, so that up and down
spin distributions can be characterized by the respective chemical
potentials $\mu _{\uparrow }^L$ and $\mu _{\downarrow }^L$,
resulting in the spin accumulation potential $\delta \mu _{s}=\left(
\mu _{\uparrow }^L-\mu _{\downarrow }^L\right) $. In the right ($R$)
wire $\delta \mu _{s}$  is much smaller,  if the spin relaxation is
faster than the influx of polarization from the left reservoir
through the ferromagnetic contact. This is satisfied when the
contact resistance is much larger than the resistance of the wire of
the length $l_s=\sqrt{D\tau _{s}} $, where $D$ is the diffusion
constant and $\tau _{s}$ is the spin relaxation rate. This is true
in many practical cases, in particular, in the systems studied in
Ref. \onlinecite{Wees,Wees2}. We, therefore, simplify our model
assuming $\mu _{\uparrow }^R=\mu _{\downarrow }^R=\mu-eV/2$   and
$\mu _{\uparrow }^L+\mu _{\downarrow }^L=2\mu + eV$, where $\mu$ is
the equilibrium chemical potential and $V$ is the charge potential
difference between two wires.

With these definitions
the electric current through the contact in the normal state is
\begin{equation}
I_n=\frac{V}{R_{c}}+\frac{\delta \mu _{s}}{2eR_{s}}\,,  \label{I}
\end{equation}%
where the inverse charge and spin resistances are given by
$R_{c}^{-1}=R_{\uparrow }^{-1}+R_{\downarrow }^{-1}$ and
$R_{s}^{-1}=R_{\uparrow }^{-1}-R_{\downarrow }^{-1}$, respectively.
In an open circuit, electro-neutrality requires $I=0$, and Eq.
(\ref{I}) gives $V=-\delta \mu _{s}R_{c}/2eR_{s}=-\delta \mu
_{s}P/2e$, where $P=R_{s}^{-1}/R_{c}^{-1}$ is the spin current
polarization of the contact. This is just the voltage induced by the
spin current trough the contact, as it has been experimentally
demonstrated in Ref. \onlinecite{Johnson2,Wees,Wees2}.

Let us now consider this situation for superconducting wires. We
assume that $2\delta \mu _{s} \ll k_{B}T_{c}$, so that the
nonequilibrium spin polarization does not cause depairing
\cite{Takahashi}. The difference between Cooper pair energies on
opposite sides of the contact is $2eV$. This potential difference
gives rise to the AC Josephson current $I_{J}$ of condensed
electrons. In addition, electro-neutrality causes an oppositely
directed current $I_{n}$ of quasi-particles, so that the total
electric current is zero. This results in  DC and AC voltage
differences between the left and right superconductors that we will
now compute.

The simplest approach to this problem is based on the assumption
that in the vicinity of the critical temperature $T_{c}-T\ll T_{c},$
the quasiparticle current remains expressed by Eq. (\ref{I}). We
will discuss in the next section in which regime this approach is
valid. Denoting the phase difference of the order parameters between
the left and right wires as $\phi $, and taking into account that
$d\phi /dt=-2eV/\hbar $, electro-neutrality $ I_{n}+I_{J}=0$ and Eq.
(\ref{I}) dictates
\begin{equation}
I_{c}\sin \phi -\frac{\hbar }{2eR_{c}}\frac{d\phi }{dt}+\frac{\delta \mu _{s}%
}{2eR_{s}}=0\,,  \label{Psi}
\end{equation}%
where $I_{c}$ is the critical Josephson current. It is easy to see
that when $I_{c}\ll \delta \mu _{s}/2eR_{s}$ the Josephson current
is dominated by harmonic oscillations with the frequency $\omega
=P\delta \mu _{s}/\hbar \equiv 2eV_{0}/\hbar $. Hence, the voltage
induced by the spin current is
\begin{equation}
V(t)=-V_{0}-I_{c}R_{c}\sin \omega t\,.  \label{V}
\end{equation}%
The DC component $-V_0$ of this voltage is exactly the same as in
the case of normal metals.  Additionally, $V(t)$ contains a term
that oscillates with a frequency determined by the DC (normal state)
contribution of the non-local signal. The magnitude of the
oscillating voltage can be estimated by noting that at temperatures
close to $T_{c}$
\begin{equation}
I_{c}=\frac{\zeta }{R_{c}}\frac{\pi \Delta ^{2}}{4ek_{B}T_{c}}\,,
\label{jc}
\end{equation}%
where $\Delta $ is the superconducting gap and $%
\zeta $ is a dimensionless coefficient that takes into account the
depairing effect inside the ferromagnetic contact layer
\cite{Efetov},  leading to exponential suppression of the Josephson
current and, consequently to small values of $\zeta $. It should be
noted that according to
Eqs. (\ref{V}) and (\ref%
{jc}) the oscillation amplitude  $I_{c}R_c$ does not
explicitly depend on the transmission coefficient, apart from the
weak dependence through the depairing factor $\zeta $.

\section{Nonequilibrium effects and collective modes}

The above analysis was based on the assumption that near the
critical temperature, the current carried by quasiparticles can be
represented by the expression in the normal state Eq. (\ref{I}),
ignoring small corrections associated with the gap in the
quasiparticle spectrum. The small gap alone, however, does not
justify this assumption. In particular, when quasiparticles are
transmitted between the left and right  wires they may not be in the
local thermal equilibrium with the respective condensates, that have
been assumed when deriving Eq.(\ref{Psi}). To take into account
nonequilibrium effects, one needs to consider time dependent
transport and relaxation of the quasiparticles. There are two
physical effects that determine kinetics of quasiparticles in the
superconducting wires. The first one is the so called
\cite{Tinkham,Pethick,Schmid} charge (or branch) imbalance of
electron and hole excitations. It is produced by quasiparticle
tunneling between superconducting electrodes, leading to a
quasiparticle distribution with a local chemical potential different
from that of the condensate. This difference relaxes during a time
much longer than the electron-phonon scattering time. Another effect
is related to condensate space-time oscillations. It dominates over
the charge imbalance relaxation when $\omega$ is large enough. We
will demonstrate that the spin injection then enables detection of
collective condensate-quasiparticle modes, Carlson-Goldman modes
\cite{Carlson} which are characterized by oppositely directed
oscillations of condensate and normal fluids. There is an important
difference with respect to the usual Josephson effect, since our
device requires no net current $I=0$. The usual Josephson effect
does not couple to Carlson-Goldman modes and is not reduced at low
temperatures, $T \rightarrow 0$. In contrast, to provide a
counterflow we need excitations that vanish at low temperatures. The
coupling to the collective modes is enabled by a spin-driven battery
effect induced by the spin injection.

Let us now detail the calculations. Assuming a small deviation from
equilibrium we employ the linearized time dependent kinetic and
Ginzburg-Landau equations in the diffusive regime \cite{Schmid},
when the elastic mean free path is much less than the
superconductor's coherence length, as well as other relevant length
scales. In this case, the isotropic quasiparticle distribution
function $f_{\sigma}(E,t)$, where $\sigma$ is the spin projection,
depends only on the energy $E$ and time $t$. Within the linear
theory the singlet condensate couples to the spin-independent part
$f(E,t)\equiv (f_{\uparrow}(E,t)+f_{\downarrow}(E,t))/2$ of the
distribution function. Therefore, after ignoring small terms
$(\delta\mu_s /k_B T)^2$, the unperturbed spin-independent
distributions takes the form of Fermi equilibrium functions
$f_0^{L/R}(E,t)$ of the left and right wires with respective
electrochemical potentials $eV/2+\mu$ and $-eV/2 + \mu$. In its
turn, the corresponding gap functions of unperturbed condensates are
$\Delta \exp (i\phi/2 -2i\mu t)$ and $\Delta \exp (-i\phi/2-2i\mu
t)$. It is easy to see that in this unperturbed state the spin
independent contribution to the quasiparticle current through the
contact is given by the first term in the right-hand side of Eq.
(\ref{I}). Taking into account above condensate functions one can
easy obtain Eqs. (\ref{Psi}) and (\ref{V}). In the perturbed state
we have $f(E,\mathbf{r},t)=f_0(E,t)+\delta f(E,\mathbf{r},t)$ (we
will skip here and below the labels $L$ and $R$). Since the
perturbation violates the electron-hole symmetry, it gives rise to a
spatially dependent potential $\varphi(\mathbf{r},t)$ near the
contact. Also, a correction to the order parameter $\delta \Delta
(\mathbf{r},t)$ appears. In order to simplify the further analysis,
we assume that $\omega \ll \Delta$ and $1/\tau _{E} \ll \Delta$,
where $1/\tau _{E}$ is the electron-phonon relaxation rate. Besides
that, the critical supercurrent $I_{c}$ is taken small enough, so
that the time dependence of all functions is dominated by harmonic
oscillations. Accordingly, we introduce the time Fourier components
$\delta f_{\omega }(\mathbf{r},E)$,
$\delta\Delta_{\omega}(\mathbf{r})$ and
$\varphi_{\omega}(\mathbf{r})$. From Refs.
\onlinecite{Schmid,Ambegaokar} it follows that $f_{\omega }$ obeys
the kinetic equation
\begin{eqnarray}
\left(-i\omega N_{1}-\widetilde{D}\nabla ^{2}+2\Delta N_{2}\right)
\delta f_{\omega }+ && \notag  \label{kinetic} \\
i\omega N_{1}f_{0}e\varphi _{\omega }-\omega N_{2}f_{0}\delta\Delta
_{\omega } &=&I_{st}\,,
\end{eqnarray}%
where $f_{0}=1/4k_{B}T\cosh ^{2}(E/2k_{B}T)$ and $I_{st}$ is the
electron-phonon scattering integral, whose explicit form can be
found in Ref. \onlinecite{Schmid,Ambegaokar}. Furthermore, $\widetilde{D}=D(N_{1}^{2}+N_{2}^{2})$, where $%
N_{1}$ and $N_{2}$ are the spectral functions:
\begin{equation}
2N_{1/2}=G_{1/2}(E+\widetilde{%
\omega }/2) \mp G_{1/2}(E-\widetilde{\omega }/2),
\end{equation}
where $G_{1}(E)=E/\sqrt{%
E^{2}-\Delta ^{2}}$ and $G_{2}(E)=i\Delta /\sqrt{E^{2}-\Delta ^{2}}$, with $%
\widetilde{\omega }=\omega +i/\tau _{E}$. In its turn, the linearized
Ginzburg-Landau equation takes the form
\begin{equation}\label{GL}
-i\omega \delta\Delta _{\omega }+\frac{8ik_B T\Delta}{\pi|\Delta
|}\int dEN_{2}\delta f_{\omega } =D\nabla ^{2}\delta\Delta _{\omega
}\,.
\end{equation}%

We will employ the above equations for the analysis of our model in
two limiting cases of weak  and strong energy relaxation versus the
Josephson frequency, $\tau _{E}\omega \ll 1$ and $\tau _{E}\omega
\gg 1$, corresponding to very different physical situations. In the
former case slow time variations of $\delta f$ may be ignored, so
that the quasiparticle kinetics is dominated by the charge imbalance
of electron and hole excitations. The deviation from the
thermodynamic equilibrium decreases with increasing distance from
the contact on the characteristic length scale $\sqrt{D\tau _{R}}$,
where $\tau _{R}=4k_{B}T_{c}\tau _{E}/\pi \Delta $ is the charge
imbalance relaxation time, that is much longer than $\tau _{E}$. In
the opposite high-frequency regime inelastic collision processes are
not important, because the quasiparticle distribution oscillates
fast. Therefore, one can neglect $1/\tau _{E}$ and $I_{st}$ in Eqs.
(\ref{kinetic},\ref{GL}). In this case, since Josephson oscillations
of the condensate take place at zero total current, they strongly
couple to Carlson-Goldman modes. Therefore, one can expect such
modes to be excited near the contact and propagate along the left
and right wires.

In both low-frequency and high-frequency regimes, using Eqs.
(\ref{kinetic},\ref{GL}) with a reduced form of $I_{st}$ from Refs.
\onlinecite{Ambegaokar} and \onlinecite{Schmid} and taking into
account the zero electric current condition, one  arrives to the
equation for the potential
\begin{equation}
\kappa ^{2}(\omega )\varphi _{\omega }=\nabla ^{2}\varphi _{\omega }\,,
\label{phi}
\end{equation}%
where $\kappa ^{2}(\omega )=1/D\tau _{R}$ at $\tau _{E}\omega \ll 1$ and
\begin{equation}
c_{s}^{2}\kappa ^{2}(\omega )=-\omega ^{2}-i\pi \omega \Delta ^{2}/
4k_{B}T  \label{kappa}
\end{equation}%
at $\tau _{E}\omega \gg 1$, where the sound velocity $c_{s}=\sqrt{2D\Delta }$%
.  Eq. (\ref{phi}) is well known. At $\tau _{E}\omega \ll 1$ it
describes the charge imbalance relaxation \cite{Schmid,Pethick},
while in the opposite limit it gives the dispersion of
Carlson-Goldman modes \cite{Schmid2}.

For our geometry, when $\kappa ^{-1}$ is much larger than the width
and thickness of the wire, $\varphi _{\omega }$ depends only on the
coordinate $x$ along the wire. Then, at $\tau _{E}\omega \ll 1$,
$\varphi _{\omega }$ exponentially decreases with increasing
distance from the contact, while at $\tau _{E}\omega \gg 1$ it shows decaying oscillations. We
assume that the left and right wires are of the same length $L$.
Since the system is symmetric with respect to $x\rightarrow -x$,
the oscillating part of the electrochemical potential is $-V_{\omega }/2+\varphi _{\omega }(x)$ at $x>0$ and $%
 V_{\omega }/2-\varphi _{\omega }(-x)$ at $x<0$, where
$V_{\omega}$ denotes the Fourier component of $V(t)$.  The solution
of Eq. (\ref{phi}) has the form
\begin{equation}
\varphi _{\omega }(x)=\alpha e^{\kappa (\omega )x}+\beta e^{-\kappa (\omega
)x}  \label{potential}
\end{equation}%
with the boundary conditions $\nabla _{x}\varphi _{\omega }(\pm L)=0$ and
\begin{equation}
\frac{V_{\omega }-2\varphi _{\omega }(0)}{R_{c}}=A\sigma \nabla _{x}\varphi
_{\omega }(0)\,,  \label{bc}
\end{equation}%
where $A$ is the wire cross-section area and $\sigma $ is the normal state
conductivity. These boundary conditions provide a zero electric current of
quasiparticles at the wire ends and the current equal to the injected one at
$x=0$. From Eqs. (\ref{phi}),(\ref{potential})-(\ref{bc}) one obtains a
periodic part of the injected current
\begin{equation}
\frac{V_{\omega }-2\varphi _{\omega }(0)}{R_{c}}=\frac{V_{\omega }}{%
R_{c}+2R_{w}}\equiv \frac{V_{\omega }}{R_{eff}(\omega )}\,,  \label{Reff}
\end{equation}%
where
\begin{equation}
R_{w}=\frac{1}{A\sigma \kappa (\omega )}\frac{1+e^{-2\kappa L}}{%
1-e^{-2\kappa L}}\,\,\,\,\,\,\,,\text{Re}(\kappa )>0\,.  \label{Rw}
\end{equation}%
Hence, the result is a renormalization of $R_c$ in Eq. (\ref{I}),
such that $R_c \rightarrow R_c+2R_w$. To find the voltage $V_{\omega
}$, the quasiparticle current (\ref{Reff}) must be equated with the
Josephson current. By this way we obtain a new expression for a time
dependent part of $V$, instead of the second term in the right-hand
side of Eq. (\ref{V}):
\begin{equation}\label{newV}
V+V_{0}=-I_{c}[\cos \omega t\text{Im}R_{eff}(\omega )+\sin \omega t\text{Re}%
R_{eff}(\omega )]
\end{equation}

\section{Discussion}

Let us analyse above results in some limiting cases.  Since
$\kappa(\omega) \rightarrow \infty$ in both cases of high
frequencies $\omega \rightarrow \infty$ and strong energy relaxation
$\tau_E \rightarrow 0$, it follows from Eq. (\ref{Rw}) that $R_w
\rightarrow 0$. We thus obtain Eq. (\ref{V}), that is an expected
result, because  in these limits a deviation from equilibrium is
small. On the other hand, the nonequilibrium effect of
quasiparticle's kinetics becomes strong when $R_c \lesssim 2R_w$.
The assumed linearization condition, however, restricts this
inequality. This condition can be expressed in the form $\zeta
|R_w/R_c| \ll 1$. Therefore, the linear theory allows $R_c \lesssim
R_w$ only at small $\zeta$. It should be noted that, according to
Eq. (\ref{Rw}), $R_w$ can be enhanced due to resonances of Josephson
oscillations with collective modes at Im$\kappa L = \pi n$, if they
are not overdamped (if Re$\kappa L \ll 1$). $R_w$ also increases at
small enough $L$, when $2|\kappa| L \ll 1$. In practice $R_w$ may be
varied in quite wide range. In Al wires from Ref. \onlinecite{Wees2}
$V_0=10^{-6}$ V, resulting in $ \omega^{-1} \simeq 0.3 \cdot 10^{-9}
$s. Since $ \omega^{-1} \sim \tau_E \sim 10^{-9} $s \cite{Wind}, a
regime intermediate between charge imbalance relaxation and
generation of Carlson-Goldman modes will be realized, with
$\kappa^{-1}$ about several $\mu$m. Therefore, strong resonances in
$R_w$ are not expected.  One can evaluate $R_w \simeq 50 \Omega$,
that is much less than $R_c=600 \Omega$.  Hence, in the considered
parameter range Eq. (\ref{V}) remains valid. In samples with higher
polarizations $P$ and at larger spin current through the contact the
Josephson oscillation frequency is expected to be large enough to
produce noticeable collective resonances
of $R_{eff}$ in Eq. (\ref%
{newV}).

The above calculations of Josephson voltage oscillations have been
restricted to $\Delta \ll T_{c}$. At the larger gap the oscillation
amplitude is expected to decrease, because less excitations are
available to compensate the supercurrent through the contact. On the
other hand, in this range one should take into account that besides
quasiparticles the spin transport through the contact can be
associated with triplet components of Cooper pair states that appear
due to spin dependent tunneling and nonequilibrium spin polarization
of superconducting wires. Further studies are needed to understand
the effect of such transport. 

In conclusion, we considered an AC Josephson effect induced by a DC
spin current through the contact whose transmittance depends on the
spin orientation of tunneling electrons. The oscillations of the
voltage across the contact at zero electric current have, in certain
parameter range, harmonic time dependence with the frequency
proportional to the spin current. The amplitude and phase of these
oscillations depend on coupled kinetics of quasiparticles and
condensate in superconducting wires. The corresponding calculations
have been performed within linearized kinetic equations at
temperature close to $T_{c}$. We predict that at the high enough
frequency the measured AC voltage will show up the resonance
structure associated with excitation of Carlson-Goldman modes.

A.G.M. gratefully acknowledges hospitality of NTNU.

\end{document}